# Scintillation properties of $SrI_2(Eu^{2+})$ for high energy astrophysical detectors: Nonproportionality as a function of temperature and at high gamma-ray energies

**R.S. Perea[a,d], A.M. Parsons[b], M. Groza[d], D. Caudel[a,d], S. Nowicki[b,c], A. Burger[a,d], K.G. Stassun[a,d], and T.E. Peterson[a,e,f]**

[a]Department of Physics and Astronomy, Vanderbilt University, 2401 Vanderbilt Place, Nashville, TN 37240

[b]NASA Goddard Space Flight Center, 8800 Greenbelt Rd., Greenbelt, MD 20771

[c]Universities Space Research Association, NASA Goddard Space Flight Center, Building 28, W174, 8800 Greenbelt Rd., Greenbelt, MD 20771

[d]Fisk University, Department of Life and Physical Sciences, 243 W.E.B. DuBois Hall, 17th Avenue North, Nashville TN 37208

[e]Institute of Imaging Science, Vanderbilt University, 1161 21st Avenue South Medical Center North, AA-1105, Nashville, TN 37232

[f]Department of Radiology and Radiological Sciences, Vanderbilt University, Nashville, TN, 37232

**Abstract**. Strontium iodide doped with europium ($SrI_2(Eu^{2+})$) is a new scintillator material being developed as an alternative to lanthanum bromide doped with cerium ($LaBr_3(Ce^{3+})$) for use in high-energy astrophysical detectors. As with all scintillators, the issue of nonproportionality is important because it affects the energy resolution of the detector. In this study, we investigate how the nonproportionality of $SrI_2(Eu^{2+})$ changes as a function of temperature 16 deg. C – 60 deg. C by heating the $SrI_2(Eu^{2+})$ scintillator separate from the photomultiplier tube. In a separate experiment, we also investigate the nonproportionality at high energies (up to 6 MeV) of $SrI_2(Eu^{2+})$ at a testing facility located at NASA Goddard Space Flight Center. We find that the nonproportionality increases nearly monotonically as the temperature of the $SrI_2(Eu^{2+})$ scintillator is increased, although there is evidence of non-monotonic behavior near 40 deg. C, perhaps due to electric charge carriers trapping in the material. We also find that within the energy range of 662keV – 6.1 MeV, the change in the nonproportionality of the $SrI_2(Eu^{2+})$ is about 1.5 - 2%.

**Keywords**: scintillator detectors, nonproportionality, gamma-ray detectors, Onsager mechanism, charge carriers, excitons

## 1 Introduction

High-energy (~1 MeV and above) detectors serve several important roles in space-based astrophysics missions, yet there are many aspects that still need to be improved upon. Specifically, detectors that may be built of materials that are low-cost, lightweight, and have the best sensitivity and energy resolution at x-ray and gamma-ray energies are the most ideal. Study of the cosmos at high energies has applications across a broad range of research in astrophysics and space science – from the gamma radiation of distant quasars and black holes, to the x-rays of solar storms, to the radioactive decay of minerals in the soil of planets and surfaces of asteroids in our solar system. Similarly, *in situ* astrobiological and geological investigations of the chemical makeup of planet and asteroid surfaces is optimally performed by probing with pulsed neutrons and measuring the scattered neutrons and gamma rays in this energy range [1]. In this manner, mineralogical assays can be performed to a depth of tens of centimeters below the surface without the need to physically drill, a process that is costly in power consumption.

For scintillators in use as gamma or x-ray detectors, the ionizing event frees an electron, creating a number of electron-hole pairs roughly proportional to the energy of the ionizing radiation. Ideally, each electron-hole pair forms an exciton, and then migrates to an activator site, where they recombine and emit a photon in the visible spectrum. These photons are then collected and converted to an electrical signal that is, ideally, proportional to the energy of the ionizing event, since the number of photons should be proportional to the ionizing energy. However, in reality, along each step of this process mitigating factors lead to a loss of excitons, photons or reduction in the output signal. [2] This resulting "nonproportionality" in turn limits the energy resolution. Indeed, while scintillators used as room temperature radiation detectors have advantages over wide bandgap semiconductors, their limiting factor at present is energy resolution. In general, the need for high energy resolution gamma spectrometers is two-fold: (i) to recognize structure in spectra (such as the shape of the positronium line) and identify closely located energy lines, and (ii) improve sensitivity, as the signal to background ratio improves with better resolution and the peak identifiability improves even when one energy line is involved[3]. For example, in the determination of the subsurface elemental composition of planets and asteroids application using conventional fast neutron activation analysis techniques, the critical advantage comes from both improved energy line identification and improved sensitivity[4].

Strontium iodide doped with europium ($SrI_2(Eu^{2+})$) is a scintillator material that in particular shows promise for high energy resolution, as a FWHM energy resolution at 662 keV has been reported as good as ~2.5% [5], close to the 2% of the semiconductor Cadmium Zinc Telluride, $CZT$[6]. An important question therefore is whether it is possible to improve the energy resolution of $SrI_2(Eu^{2+})$ to the point that it is even more competitive with CZT. This would lead to a scintillator with a resolution comparable to or better than that of a semiconductor, while less expensive to manufacture and with the capacity to grow much larger single crystals and therefore build more efficient detectors compared to those fabricated from semiconductors [7].

In order to improve the energy resolution of $SrI_2(Eu^{2+})$ requires characterizing and better understanding the nonproportionality of the material's light yield. Indeed, some calculations show that, should nonproportionality be minimized, $SrI_2(Eu^{2+})$ would achieve a fundamental Poisson limit of 1.5% energy resolution [8].

Nonproportionality is a well-known characteristic of scintillators [9-11] and arises because the incoming photon can deposit its full energy in a variety of ways (e.g. Compton Scattering, Auger electrons, etc.) in a "cascade" process, which causes variations in the amount of light produced within the detector [10], and the light-yield responses to the "individual divisions of energy" are not proportional to that energy[10]. It is thought that the underlying cause of nonproportionality is connected to the details of transportation of electric carriers (electrons, holes and excitons) within the scintillator [9, 10, 12, 13]. The link between the transportation of charge carriers and the amount of light that a scintillator produces (the luminosity of the scintillator) was originally quantified by Birk's equation [2]. Attempts at modeling nonproportionality generally begin with this form, or an empirically modified form of Birk's equation [2]. One of these modified forms, that used in Payne et al., incorporates the "Onsager Mechanism" into their model for light yield of a scintillator [13]. The Onsager mechanism is a recombination rate that depends on the "Onsager radius" (separation) where an electron and hole will no longer recombine; this is because the Onsager radius is the distance where the Coulombic and thermal energies are equal [13-15]. While in the Payne et al. study, it is implied that perhaps the Onsager radius is connected with a term that pertains to the intrinsic properties of the crystal (and is verified upon comparison to experimental data), we infer that it is possible that there is also a connection between the Onsager radius and nonproportionality. This is further detailed in the discussion section.
To make progress in the effort to improve the nonproportionality and therefore the energy resolution of scintillators as a new cost-effective solution for high-energy astrophysical detectors, it is necessary to characterize how factors such as operating temperature and incoming photon energy affect the nonproportionality. Therefore, two studies with $SrI_2(Eu^{2+})$ are presented in this paper. One is a study of nonproportionality as a function of temperature. The other is a study of nonproportionality as a function of energy at high energies.

The response of $SrI_2(Eu^{2+})$ as a function of temperature has been investigated previously by Lam et al., and Alekhin et al. (both studies are specific to $SrI_2$, doped and un-doped, with $Eu^{2+}$ [16, 17]), and Boatner et al., who investigated a variety of scintillators up to very high (400 deg. C) temperatures[18]. In the Lam et al. study, they focused on temperatures from 295 K down to 5 K, while the Alekhin et al. study went from 80 K to 600 K (for $SrI_2(Eu^{2+})$) at low energies (up to 1 keV). Our study differs from the above studies in that it focuses on the nonproportionality of $SrI_2(Eu^{2+})$ from 16 deg. C to 60 deg. C using gamma-ray sources of energy ranging from 81 keV to 1275 keV. The focus of the study presented here is the shift of the photopeak with increasing temperature and its implication to how nonproportionality changes as a function of temperature (the Boatner et al. study does not focus on nonproportionality, only on the shift of the photopeak) at a higher temperature range than Lam et al. study and at higher energy ranges than Alekhin et al. Information on the nonproportionality in this energy and temperature range may add information to the underlying physical cause(s) of nonproportionality (such as Onsager Mechanism, trapping etc.)

The high energy study is unique in that this is the first study of $SrI_2(Eu^{2+})$ at such high energies (to our knowledge). The reason for this experiment was to evaluate the $SrI_2(Eu^{2+})$ detector as a possible gamma-ray alternative to the $LaBr_3(Ce^{3+})$ currently used in the Probing in-situ with Neutrons and Gamma-rays Instrument [19]. An interest in an alternative to $LaBr_3(Ce^{3+})$ comes from the need to have a scintillator without any self-activity. $LaBr_3(Ce^{3+})$ contains selfactivity due to the $^{138}La$ [20] creating intrinsic photopeaks in the gamma-ray spectrum that increases the background noise, and can further interfere with composition analysis if the particular element has a characteristic gamma-ray near any of the intrinsic ones coming from the crystal itself.

Both experiments bring new regimes to the previous studies mentioned above. Studies of the electron response such as [5, 13, 16] (to name a few) based on Compton scattering are useful and simpler for probing the effect of nonproportionality on scintillator energy resolution, since the ionization avoids the cascade process that occurs with gamma-ray interactions in inner shell electrons. Studies of gamma photon response and its temperature variation, while more complicated due to the cascade mentioned above are directly relevant to high-energy (6 MeV) gamma spectroscopy applications. Experimental set-up and results for these studies are shown in section 2, followed by the discussion in section 3, and conclusions in section 4.

## 2. Experiments
*2.1. Nonproportionality as a function of Temperature:*
*2.1.1 Set-up*

The experimental set-up consists of (in order from top to bottom on Figure 1): an outer aluminum sleeve, Teflon, a copper heater, a 0.6 cm$^3$ SrI$_2$(Eu$^{2+}$) crystal encapsulated in an aluminum can, a hollow quartz optical rod (101 mm in length) inside a brass cylindrical core surrounded by copper tubing, and a Hamamatsu photomultiplier tube (PMT) (Ultra-Bialkali model R6231-100). The crystal was grown and encapsulated at Fisk University with a FWHM resolution of 3.9 % at 662 keV when tested directly coupled to the PMT (without the light guide). While use of the light guide causes loss of scintillation light, resulting in degradation of the resolution, its use here is to separate the PMT from the heat being applied to the crystal (allowing us to keep the PMT at room temperature). The inset in Figure 1 shows the assembly of the top portion (everything except the PMT).

The method for acquiring data was done using the following three sources (energies ranging from 81 keV to 1275 keV): $^{133}$Ba, $^{137}$Cs, and $^{22}$Na. The selected temperatures used were 16 deg. C, 30 deg. C, 40 deg. C, 50 deg. C, and 60 deg. C. Counts from each source were collected for five consecutive runs at each temperature with the exception of $^{22}$Na at 40 deg. C, and at 60 deg. C. In the case of 40 deg. C, there are only two runs recorded, and in the case of 60 deg. C there are only four. Each spectrum of the $^{133}$Ba and $^{137}$Cs sources where acquired for 300 seconds, while to achieve good statistics the $^{22}$Na spectra were recorded for 900 seconds.

*2.1.2. Results*

Figure 2 shows the full spectra of the final run at each temperature of the $^{22}$Na source. As temperature increases, the 511 keV and the 1275 keV photopeaks shift to lower channels. Only the $^{22}$Na source is shown (for brevity), as the same behavior of the photopeak shifting to lower channels with increasing temperature is present for all the sources in all sets of spectra. All values are tabulated in Table 1. The centroid values shown in Table 1 are the weighted averages of all the runs for that energy and temperature.

The fitting and averaging routine that was used was written in Python, with the fitting routine making use of the Python function called curve_fit() [21]. All the spectra are fitted with a Gaussian plus a polynomial (for the background continuum). To obtain the weighted average of the data, another function of Python, called numpy.average() [22] was used.

Also, because of the light guide use, the resolution is reduced to around 9.6%. This reduction in energy resolution and relative closeness of the 276 keV and 302 keV photopeaks of the $^{133}$Ba source are unresolved, along with the 356 keV and 383 keV peaks. For this reason, of the $^{133}$Ba source, we only analyze the 81 keV photopeak along with the 511 keV and 1275 keV photopeaks of $^{22}$Na, and the 662 keV photopeak of $^{137}$Cs (which have no other source peaks nearby and are less affected) for the peak shifts and nonproportionality calculations.

To quantify the amount of shift in the photopeak, we performed a percent decrease calculation using 16 deg. C as the reference temperature:

$$Relative\ Shift\ (in\ \%) = \frac{ADC\ Channel_{Temperature} - ADC\ Channel_{16 deg.C}}{ADC\ Channel_{16\ deg.\ C}} * 100\% \quad (1)$$

$$\sigma_{Relative\ Shift\ (in\ \%)} = \left[\left(\frac{1}{ADC\ Channel_{16\ deg.\ C}}\right)^2 * \sigma^2_{ADC\ Channel\ Temperature} + \left(\frac{ADC\ Channel_{Temperature}}{ADC\ Channel_{16\ deg.\ C}^2}\right)^2 * \sigma^2_{ADC\ Channel_{16\ deg.\ C}}\right]^{1/2} * 100\% \quad (2)$$

The results are presented in Figure 3 and Table 2. The percent decrease between energies, at each temperature (with the exception of the the 511 keV and 1275 keV peaks of $^{22}$Na at 40 deg. C which are addressed in the discussion section) is 4%-6%. This decreasing trend seen at all energies can be linearly fit. A possible explanation for this seemingly linear behavior is also addressed in the discussion section.

Lastly, Figure 4 shows the nonproportionality as a function of temperature. This is calculated by the following equations:

$$Nonproportionality = \frac{ADC\ Ch\#_{Photopeak} * \frac{662\ keV}{ADC\ Ch\#_{662\ keV\ Photopeak}}}{Energy_{Photopeak}} \quad (3)$$

$$\sigma_{Nonproportionality} = \left\{\frac{662\ keV}{Energy_{Photopeak}} * \left[\left(\frac{1}{ADC\ Ch\#_{662\ keV\ Photopeak}}\right)^2 * \sigma^2_{ADC\ Ch\#_{Photopeak}} + \left(\frac{ADC\ Ch\#_{Photopeak}}{ADC\ Ch\#^2_{662\ keV\ Photopeak}}\right)^2 * \sigma^2_{662\ keV\ Photopeak}\right]^{1/2}\right\} \quad (4)$$

For all energy sources, the nonproportionality increases with the rise in temperature. This increase is about 6% overall. Tabulated values can be found in Table 3.

*2.2. Nonproportionality at High Energies:*
*2.2.1 Setup*

As mentioned in the introduction the second experiment was performed at the 6 MeV Gamma-ray Facility at NASA Goddard Space Flight Center (GSFC) with Dr. Ann Parsons and Dr. Suzanne Nowicki[23]. The set-up for producing gamma-rays is shown in Figure 5. Piping surrounds the granite monument; starting from the left of the picture, the piping is wrapped in a helical fashion; this is where the pulsed neutron generator is placed. The piping goes across the top of the monument, to a water tank on the right side of the picture, then comes back across (this is depicted in the schematic also included in Figure 5). When the neutron generator is turned on, the neutrons excite the oxygen via the $^{16}O(n,p)^{16}N$ reaction. By the time the water has reached the tank, the $^{16}N$ de-excites releasing gamma-rays of 2.7, 6.1, and 7.1 MeV [24] (with probabilities of 0.82%, 67%, and 4.9% respectively [25]).

Also shown in Figure 5 is the placement of the $SrI_2(Eu^{2+})$ detector. The $SrI_2(Eu^{2+})$ detector was loaned to us from Lawrence Livermore Laboratory, and contains a 1 in by 1 in crystal (2.95% (19.5 keV) resolution at 662 keV) grown by Radiation Monitoring Devices. For a crystal of this size, we expect about 1% - 2% efficiency at 6 MeV. For the experiment the $SrI_2(Eu^{2+})$ detector was grounded (with foil), and also wrapped in a foil "sack" to reduce interference from radio waves (used by other groups performing radar ranging at the NASA site). The placement of the $SrI_2(Eu^{2+})$ detector was about 20.5 (2) cm, and acquisition time was 5 hours.

*2.2.2. Results*

Figure 6 shows the resultant spectra of $SrI_2(Eu^{2+})$ of the 6 MeV experiment. A $^{137}Cs$ source was placed near the detector as a calibration source. Lastly, the nonproportionally of $SrI_2(Eu^{2+})$ over the range of 662 keV – 6.1 MeV was calculated and is shown in Figure 7 and in Table 4.

**3. Discussion**
*3.1: Nonproportionality as a Function of Temperature*

In Sec. 2.1 we presented the results of the nonproportionality of $SrI_2(Eu^{2+})$ as a function of temperature, see Figure 4. In Figure 4, the nonproportionality increases as the temperature increases, with the crystal most nonproportional at 60 deg. C. This degradation is in agreement with the Alekhin et al. study (who found that the nonproportionality of their x-ray response became greater as they increased the temperature from 295 K to 600 K) [16]. In comparison to the Lam, et al. study, our higher temperature measurements seem to still be in agreement with the lower temperature data [17]. Interestingly, those authors see a 5% degradation in nonproportionality as they decreased their temperatures from 295K down to 5K with their lower energy range [17] of 31 keV - 41 keV. If we do the same comparison, we see an ~6% degradation in nonproportionality for our higher energy range of 81 keV - 1275 keV as we increase the temperature from 16 deg. C – 60 deg. C. From this comparison, it is clear that nonproportionality is indeed temperature dependent, however, whether the cause at lower temperature vs. higher temperatures is due to the same mechanism cannot be verified with this study.

To begin to understand the possible physical causes of these trends in Figure 4, we observe in Figures 2-3 and Tables 2-3, that there is a decrease in the amount of scintillation photons. Within the scintillator, there are many processes that can keep a scintillation photon from being produced (or delayed), such as: radiation-less transitions to the ground activator states[2], forbidden transitions in the activator states[2], exciton- exciton annihilation[13], trapping[26], and the Onsager mechanism[13] (mentioned previously) to name a few.

The Onsager mechanism may provide a viable explanation for our results, as we see a decrease of scintillation photons with increasing temperature. In the Payne et al., 2009[13] study, the Onsager mechanism, which states that the electron and hole need to be within a particular radius in order to recombine (from a balance of Columbic and thermal energies [13]), has an inverse relationship between that radius and temperature. This inverse relationship is seen in Figure 3, where our data points have been fit to a linear equation for each energy. If in our study, the cause of less scintillation photons is due to the Onsager mechanism, then as we increased the temperature, the Onsager radius would decrease, making recombination more difficult with increasing temperature. Since the temperature was increased by the same fixed amount; the shift in channel number was decreased by the same amount.

Another possibility for the seemingly proportional behavior could be due to the mobilities of the electrons and holes. As the incoming photon ionizes the scintillator, the effects of electrostatic forces and thermal diffusion cause the charge carriers to separate. If the mobility of the charge carriers is low (or one of the charge carrier's mobility is much greater than the other), the scintillator behaves nonproportionally [9]. However, if both mobilities are high, then the scintillator behaves like a proportional crystal [9]. Currently, there is no method to measure the mobility of charge carriers within a scintillator directly (as is done with semi-conductors). If a method to accurately measure these mobilities is developed, then diffusion models such as the one mentioned here would help in interpreting these results. If the mobilities are indeed high (and perhaps more so at higher temperatures), then the condition imposed by the Onsager radius is never met for recombination.

Finally, the cause of the drop in our data points at 40 deg. C with the 511 keV and 1275 keV peaks of $^{22}$Na in Figure 4 is not clear. There are indications of potential problems with the data run that produced the $^{22}$Na measurements at 40 deg. C (see Table 1 where only 2 measurements were obtained instead of 5 for most of the other runs).

*3.2: Nonproportionality at High Energies*

Over the energy range of 662 keV – 6.1 MeV, the SrI$_2$(Eu$^{2+}$) detector experiences only a small change in nonproportionality (1.5 - 2%). Previous measurements with LaBr$_3$(Ce$^{3+}$) have shown a similarly small change in nonproportionality of at most 2% over the energy range 30 keV – 6 MeV [27]. Therefore SrI$_2$(Eu$^{2+}$) appears to have a similarly good proportionality but without the detrimental effects of self-activity.

## 4. Conclusions

In this paper, we report findings from two different experiments on the nonproportionality of SrI$_2$(Eu$^{2+}$), motivated by the need to develop effective high-energy detectors with comparable energy resolutions to semi-conductors for astrophysics applications and future space missions. One study was focused on the change in nonproportionality as a function of temperature; the other was focused on the nonproportionality at high energies.

In the temperature variation study, we found that the photopeak of the SrI$_2$(Eu$^{2+}$) crystal shifts to lower channels with increasing temperature, and that the nonproportionality increases (up to 6%) with increasing temperature. This could be due to the charge carriers being at a distance equal to or larger than the Onsager radius as their kinetic energy is increased with the higher temperature. This increase in nonproportionality in temperature is in agreement with previous studies [16], and also seems to be a trend when the temperature is decreased [17]; whether this is caused by the same mechanism is left for further investigation (and perhaps can be answered with further investigations of the mobilities of the charge carriers in the scintillator).

These results show that increasing temperature increases the nonproportionality, further investigation is needed to tie the specific cause (Onsager Mechanism, trapping and/or mobilities) and whether this cause is the same for high or low temperatures. Understanding this cause could give guidance into how to improve the growth (or perhaps correctly compensate for the effect in an already fabricated detector) so that this effect can be mitigated and energy resolution may be improved.

In the high energy study, we found that $SrI_2(Eu^{2+})$ has a nonproportionality of about 2% at 6 MeV. Within the energy range of 662 keV – 6.1 MeV, the change in the nonproportionality of the $SrI_2(Eu^{2+})$ is about 1.5 - 2%.

This is the first study of $SrI_2(Eu^{2+})$ at this energy regime, and these preliminary findings seem to be promising for applications in astronomy. If potential improvement in the nonproportionality of $SrI_2(Eu^{2+})$ is achieved, then $SrI_2(Eu^{2+})$ can become a viable detector with low background and no selfactivity. This would enable lower cost high-energy detectors deployable both in space-based platforms and in less temperature-controlled environments such as planetary surface landers performing compositional analyses.

**Acknowledgments**

We would like to thank the two anonymous reviewers for their constructive criticism that helped improve this paper.

One of the authors (RP) would like to acknowledge the financial support of the following: "Graduate Opportunities at Fisk in Astronomy and Astrophysics Research (GO-FAAR)" and the IGERT DGE-1069091 "Neutron Scattering for the Science and Engineering of the 21st Century", support provided by the NSF grants, AST-0849736 (PAARE) as well as the support through NASA's 2012 Lunar and Planetary Science Academy at the Goddard Space Flight Center, and University of Maryland's 2013 Center for Research and Space Exploration in Space Science and Technology (CRESST).

We would like to thank Nereine Cherepy, Steve Payne, Patrick Beck, Scott Fisher and Peter Thelin at LLNL for providing us with the packaged $SrI_2(Eu^{2+})$ detector and their guidance for the 6 MeV experiment. We would also like to thank Rastgo Hawrami and Kanai Shah at RMD Inc. for providing us the large $SrI_2(Eu^{2+})$ crystal that was used in fabricating the detector.

One of the authors (RP) would also like to acknowledge the Fisk-Vanderbilt Masters-to-PhD Bridge Program and the Bridge Postdoc (Rudolfo Montez) for the support of this paper and in creation of the Python scripts utilized for the temperature data analysis.8

**R.S. Perea** is a graduate student at Vanderbilt University in the Physics and Astronomy Department. She received her B.S. in Engineering Physics (Mechanical Engineering option) in 2007 from New Mexico State University (NMSU) an M.S. in Physics (2010, NMSU) and an MA in Physics (2013 Fisk University). Through the Fisk-Vanderbilt Masters-to-the-PhD program, she has obtained experience with both scintillators/semiconductors and two internships at the NASA/GSFC. This combination has prepared her in detector/instrument design for astrophysics.

Biographies of other authors not available.

**Table 1.** Summary of averaged centroid values. Below are the weighted average centroid values (in ADC channel) for each temperature.

| Energy (keV) | 16 deg. C Mean (error) | 30 deg. C Mean (error) | 40 deg. C Mean (error) |
|---|---|---|---|
| 81 | 61.524 (0.023) | 58.157 (0.061) | 55.803 (0.041) |
| 511 | 375.789 (0.613) | 356.011 (0.102) | 333.547 (0.781) |
| 662 | 487.596 (0.131) | 459.730 (0.082) | 439.375 (0.016) |
| 1275 | 925.454 (1.347) | 876.264 (0.389) | 821.248 (1.414) |
| Energy (keV) | 50 deg. C Mean (error) | 60 deg. C Mean (error) | |
| 81 | 53.025 (0.035) | 50.540 (0.045) | |
| 511 | 323.848 (0.074) | 307.630 (0.057) | |
| 662 | 417.655 (0.058) | 396.787 (0.120) | |
| 1275 | 797.003 (0.696) | 757.137 (2.000) | |

Table 2. Relative shift of photopeak in $SrI_2(Eu^{2+})$. Tabulated values for the shift in photopeak of $SrI_2(Eu^{2+})$ as a function of temperature. Values correspond to Figure 3.

| Energy (keV) | Between 30 deg. C and 16 deg. C Relative Shift (in %) (Error) | Between 40 deg. C and 16 deg. C Relative Shift (in %) (Error) |
|---|---|---|
| 81 | -5.468 (0.105) | -9.298 (0.075) |
| 511 | -5.262 (0.157) | -11.240 (0.253) |
| 662 | -5.714 (0.030) | -9.889 (0.024) |
| 1275 | -5.315 (0.144) | -11.259 (0.200) |
| Energy (keV) | Between 50 deg. C and 16 deg. C Relative Shift (in %) (Error) | Between 60 deg. C and 16 deg. C Relative Shift (in%) (Error) |
| 81 | -13.813 (0.221) | -17.852 (0.079) |
| 511 | -13.821 (0.142) | -18.137 (0.134) |
| 662 | -14.341 (0.026) | -22.605 (0.033) |
| 1275 | -13.879 (0.187) | -18.624 (0.625) |

**Table 3.** Nonproportionality of $SrI_2(Eu^{2+})$. Tabulated values of the nonproportionality vs. temperature as shown in Figure 4.

| Energy (keV) | Nonproportionality (error) 16 deg. C | Nonproportionality (error) 30 deg. C |
|---|---|---|
| 81 | 1.031 (4.768E-04) | 1.034 (1.091E-03) |
| 511 | 0.998 (1.650E-03) | 1.003 (3.384E-04) |
| 662 | 1.000 (3.788E-04) | 1.000 (2.529E-04) |
| 1275 | 0.985 (1.458E-03) | 0.989 (4.743E-04) |
| Energy (keV) | Nonproportionality (error) 40 deg. C | Nonproportionality (error) 50 deg. C |
| 81 | 1.038 (7.684E-04) | 1.038 (6.991E-04) |
| 511 | 0.983 (2.302E-03) | 1.005 (2.696E-04) |
| 662 | 1.000 (5.048E-05) | 1.000 (1.961E-04) |
| 1275 | 0.970 (1.672E-03) | 0.991 (8.762E-04) |
| Energy (keV) | Nonproportionality (error) 60 deg. C | |
| 81 | 1.041 (9.853E-04) | |
| 511 | 1.004 (3.570E-04) | |
| 662 | 1.000 (4.286E-04) | |
| 1275 | 0.991 (2.634E-03) | |

**Table 4.** Nonproportionality values of $SrI_2(Eu^{2+})$ at energies up to 6 MeV (normalized to 662 keV), along with the average nonproportionality (data point in red in Figure 7). Energies of the $^{16}O(n,p)^{16}N$ reaction can be found at the National Nuclear Data website (see reference 24).

| Energy (keV) | NonProportionality (error) | Energy Resolution in % (error) |
|---|---|---|
| 662 | 1.000 (8.746E-05) | 4.842 (0.012) |
| 5107 | 0.987 (5.755E-04) | 3.955 (0.115) |
| 5618 | 0.994 (8.751E-04) | 4.689 (0.179) |
| 6129 | 0.979 (1.272E-03) | 2.989 (0.278) |
| Average Energy | Average Nonproportionality | |
| 5618 | 0.987 (9.074E-04) | |

## **Figure Caption List**

**Figure 1.** Cross-sectional view of temperature experiment. Figure shows components of temperature study set-up.

**Figure 2.** $SrI_2(Eu^{2+})$ $^{22}Na$ spectra as a function of temperature. Figure shows the final run for each temperature with the $^{22}Na$ source. Notice that the photopeaks shift to lower channels with increasing temperature (please see color version online).

**Figure 3.** Figure is a graphical representation of the relative shift of the centroid value at different energies, from a reference temperature of 16 deg. C. Values can be found in Table 2. Each energy has a corresponding linear fit to the data points (please see color version online).

**Figure 4.** Nonproportionality of $SrI_2(Eu^{2+})$ vs. temperature. Figure shows the nonproportionality vs. temperature, with data normalized to 662 keV at each temperature. The data points at 511 keV and 1275 keV are addressed in section 3.1.

**Figure 5.** 6 MeV gamma-ray facility set-up. Picture of the 6 MeV gamma-ray facility at NASA GSFC. Neutron pulsed generator is on the left, and the water tank is on the right. Gamma-rays are produced via the $^{16}O(n,p)^{16}N$ reaction. The schematic shows placement of neutron generator, water tank, and $SrI_2(Eu^{2+})$ detector for the experiment set-up.

**Figure 6.** Spectrum of $^{137}Cs$, and 6 MeV gamma-ray sources obtained with $SrI_2(Eu^{2+})$. Peaks seen in the spectra are the $^{137}Cs$ photopeak, and the $^{16}O$ photopeak, see reference 24, ($^{16}O$ escape peaks are also seen). Inset shows fits (Gaussian + polynomial) to these photopeaks (please see color version online).

**Figure 7.** Nonproportionality vs. Energy of $SrI_2(Eu^{2+})$, normalized to 662 keV. The red square (color version online) represents the average nonproportionality in the high energy range. The error bars of this point in the x-axis represent the range of energies measured, while the error bars in the y-axis represent the true error in the nonproportionality. Values are shown in Table 4.

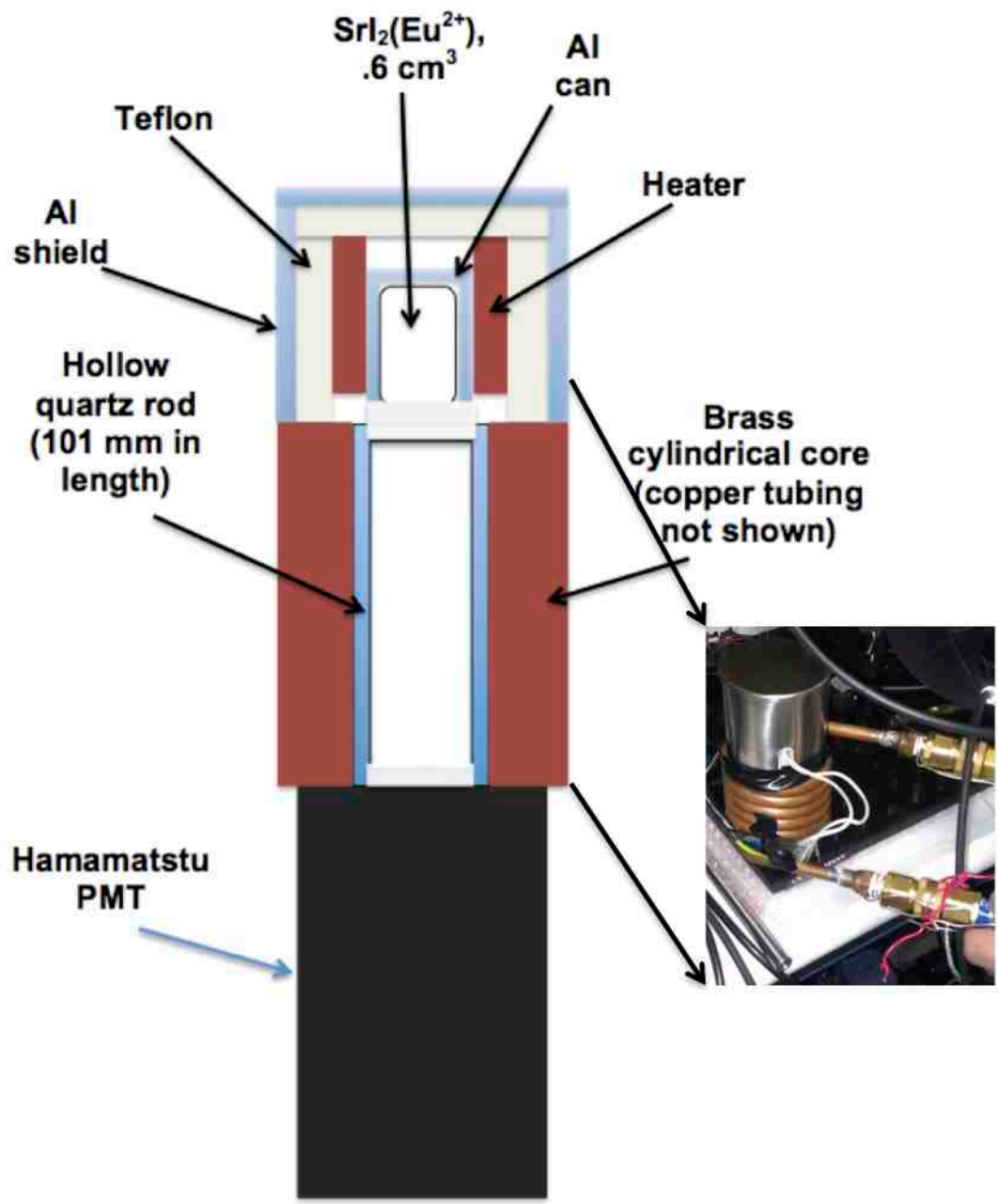

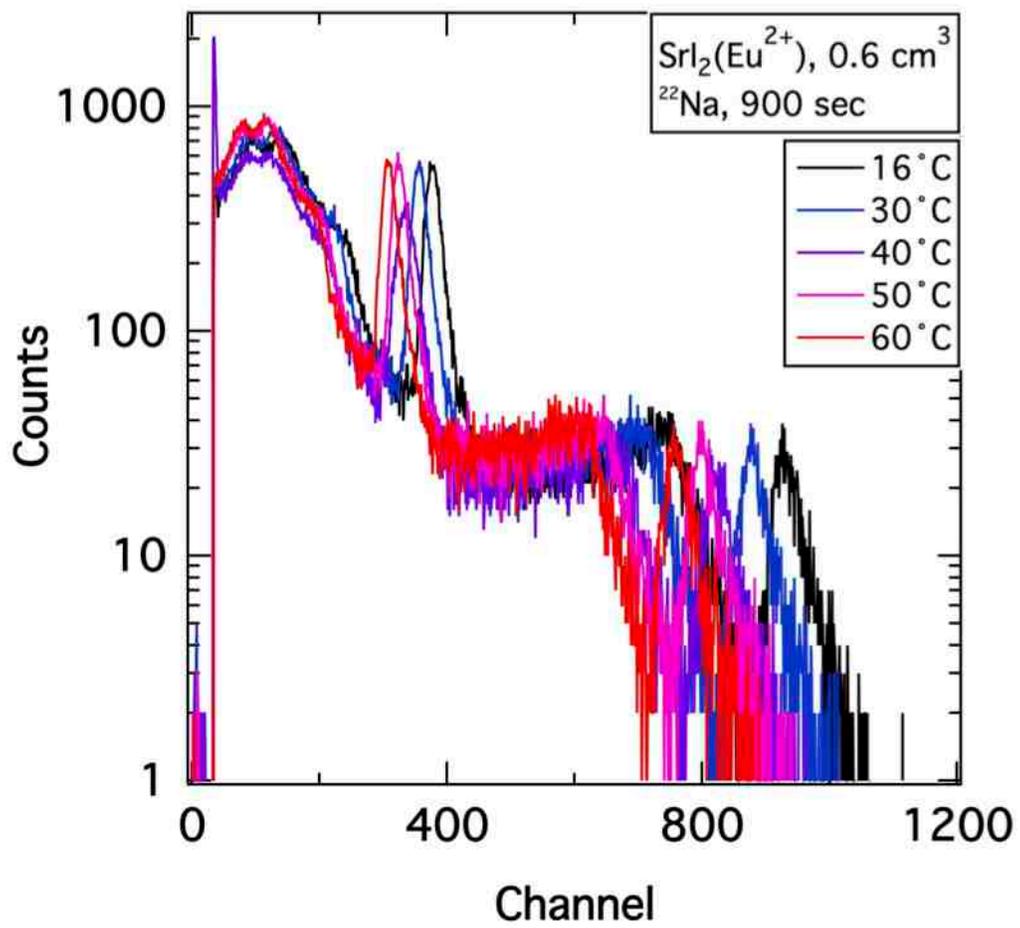

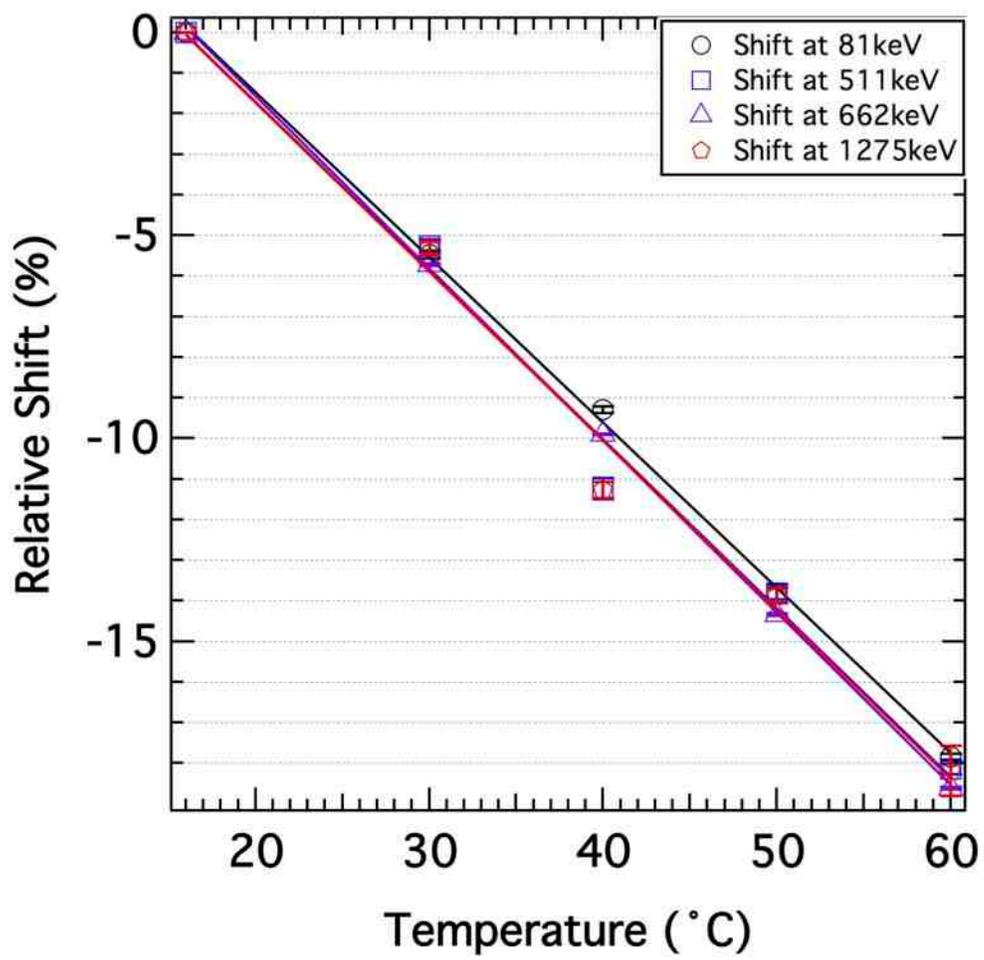

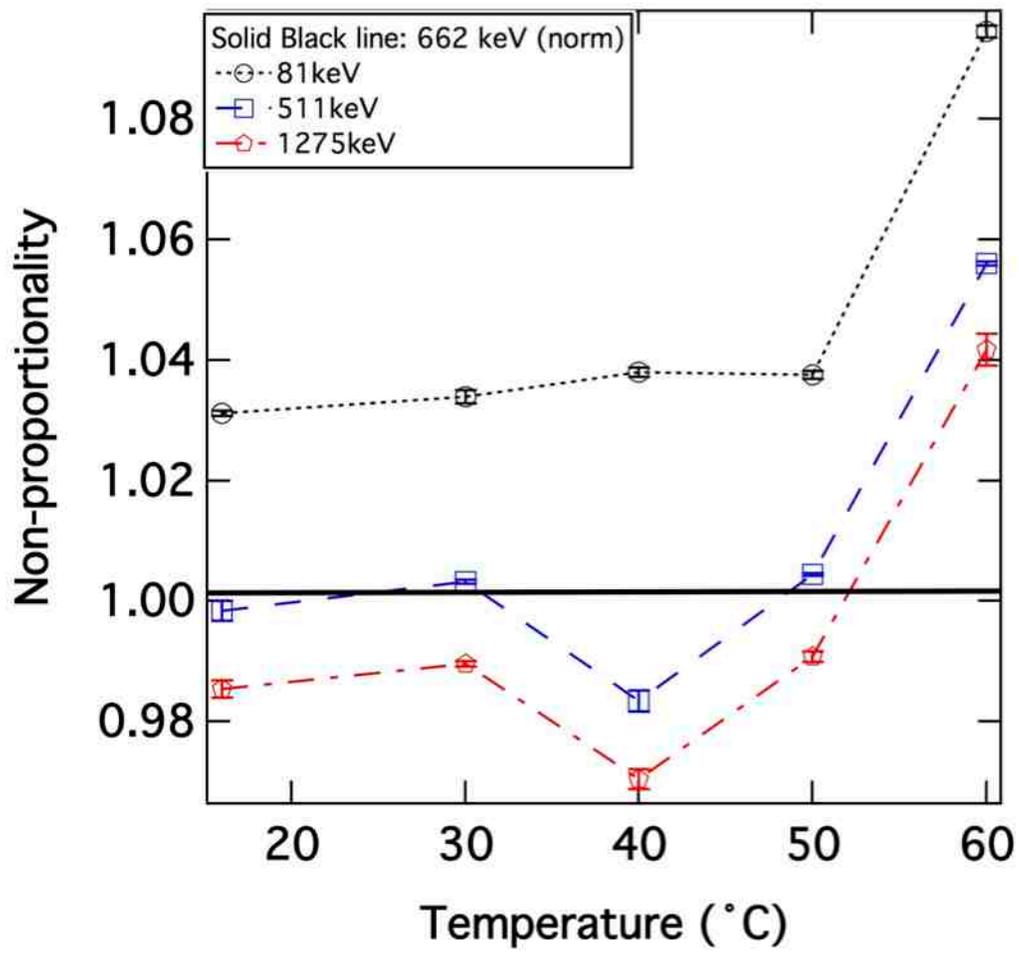

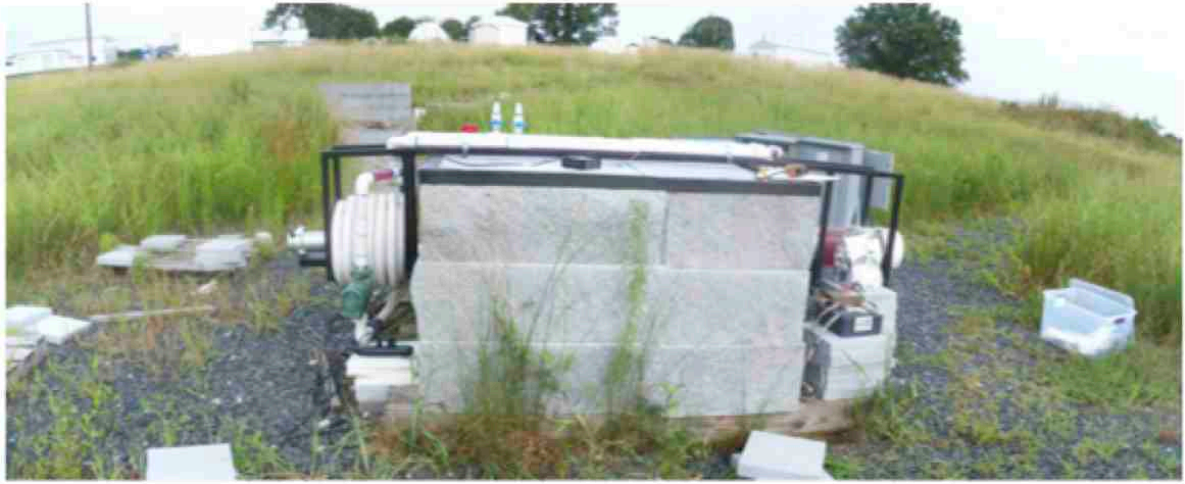

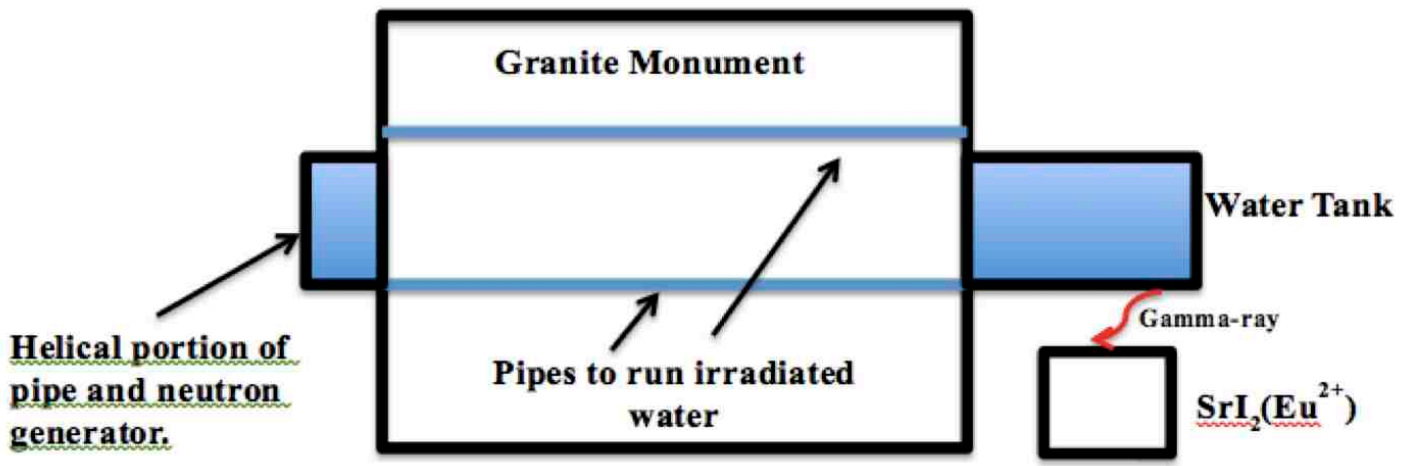

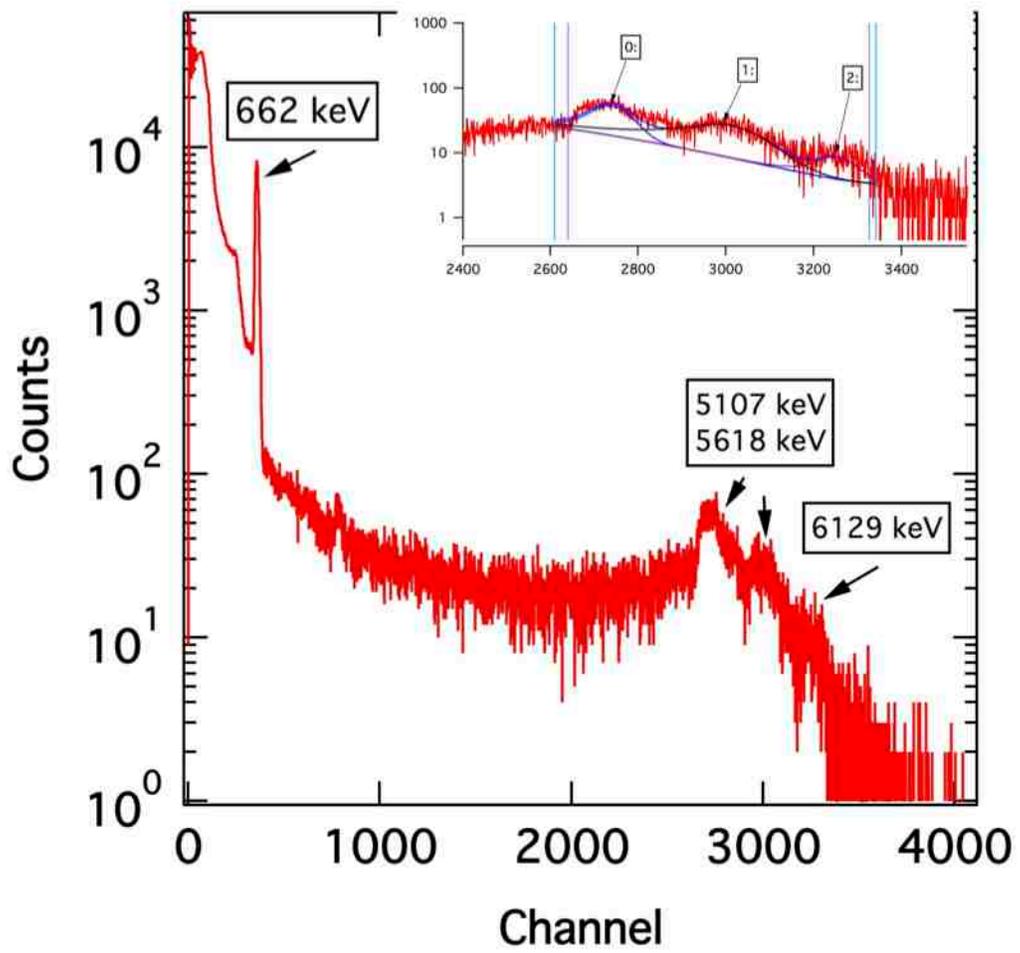

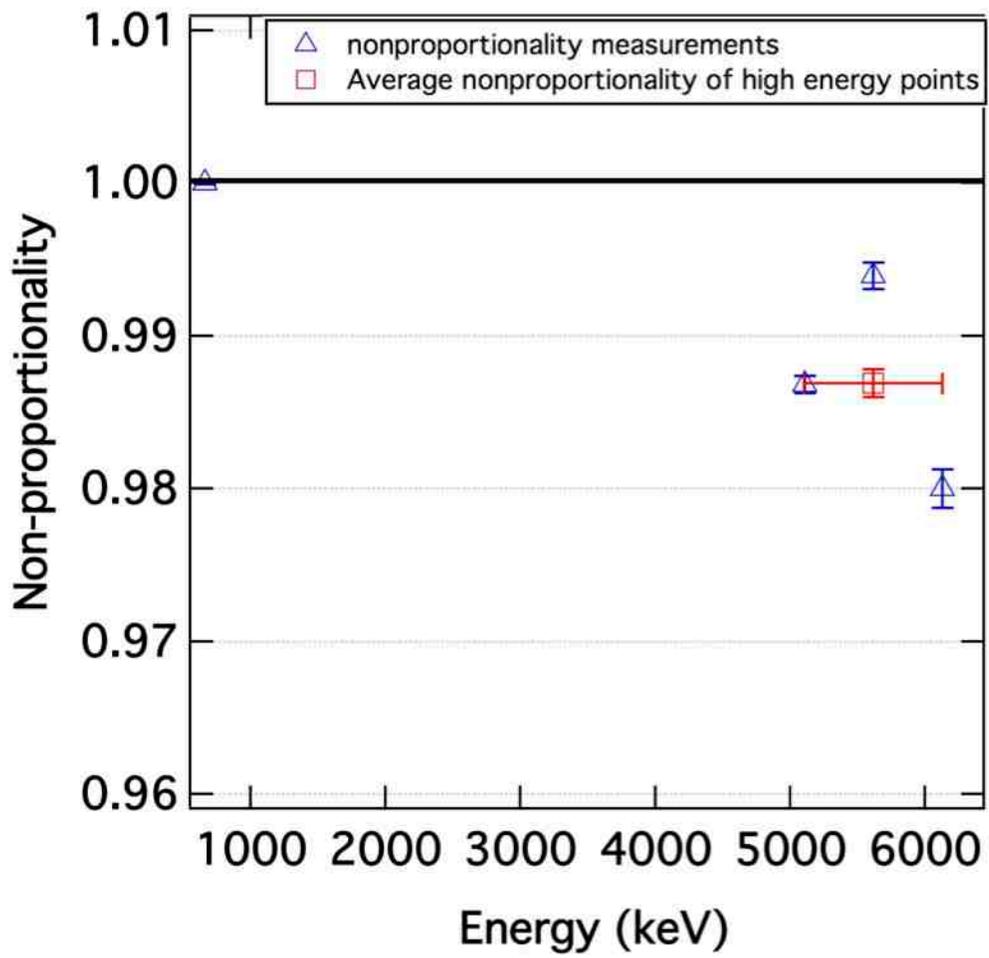